\documentclass{ws-ijgmmp}

\begin{document}

\markboth{Partha Sarathi Debnath, Chanchal Mondal, B. C. Paul}
{Dynamical analysis of bulk viscous cosmological model  in $F(T)$  gravity}

\catchline{}{}{}{}{}
%
%%%%%%%%%%%%%%%%%%%%%%%%%%%%%%%%%%%%%%%%%%%%%%%%%%%%%%%%%%%%%%%%%%%%

\title{Dynamical analysis of bulk viscous cosmological model  in $F(T)$  gravity}

\author{Partha Sarathi Debnath}

\address{Dept. of Physics, A. P.C. Roy Govt. College 
Matigara, Siliguri, W. B., Pin-734010, India. \&\\
Dept. of Physics, Alipurduar Govt. Engineering and Management College, Alipurduar, W.B. Pin 736206, India\\
\email{partha.debnath@associates.iucaa.in} } 
\author{Chanchal Mondal, B. C. Paul}

\address{Department of Physics, University of North Bengal, Rajarammohanpur, \\
Siliguri, West Bengal, Pin-734013, India.\\
\email{bcpaul@associates.iucaa.in} } 

\maketitle

\begin{history}
\received{(Day Month Year)}
\revised{(Day Month Year)}
\end{history}

\begin{abstract}
Bulk viscous cosmological models is presented in the teleparallel ($F(T)$, where $T$ denotes torsion) gravity. In the teleparallel gravity, the Lagrangian of the gravitational action contains a general function $F(T)= T+ f(T)=(1+ \gamma) T+\alpha (-T)^n$, where $\gamma$, $n$ and $\alpha$ are dimensional constants.  Cosmological solutions in Eckart theory and Truncated Israel Stewart theory are talked about in $F(T)=(1+ \gamma) T+\alpha (-T)^n$ gravity form which is one of the most generalized gravity form in the torsional gravity. The substantial and geometrical prospective of the cosmological models in Eckart theory and Truncated Israel Stewart theory in $F(T)$  gravity are deliberated for flat Friedmann-Robertson-Walker space time. Dynamical analysis of the fixed points of exponential expansion in $F(T)$ with bulk viscous cosmological models are studied here.  The characteristics of the various cosmological parameters such as bulk viscous pressure, energy density, scale factor, Hubble parameter and entropy evolution are studied in Power law and Exponential models. Stability analysis of the exponential models are also argued with cosmic growth in the relevant theories by using directional plots. The cosmological models are supported by observations results.
\end{abstract}

\section{Introduction}
The current cosmic accelerating phase of evolution of the universe is proposed by modern cosmic data \cite{Riess}.  General theory of gravity (GTR) is a very active theory to demonstrate most of the gravitational phenomena for the development of the universe. However,   standard cosmological models with perfect fluids fail to build up the late-time acceleration of the universe. The late-time acceleration \cite{Padma} dilemmas along with the dark matter difficulty are the most demanding assignments to the cosmologist.

The late-time acceleration is credited to a negative pressure fluid called as dark energy \cite{Cope}, but what is the dark energy for the moment we have absolutely no idea.  The PLANCK Collaboration 2015 estimates that the  dark energy which plays a significant role to compel the acceleration, as  consisting of  $\sim $ 69.4$\%$ of the total energy, dark matter contributes $\sim $ 25.8 $\%$ and baryonic matter consists of $\sim $ 4.8$\%$. The theoretical suggestions \cite{Nojiri11} came up to recognize the correct fundamental nature of the dark energy. Phenomenological models \cite{starobinsky,mukherjee}  appear by modifying gravitational sector and/or matter sector of Einstein’s field equation to study geometrical and substantial features of the universe. The modified theories of gravity with suitable curvature alteration are also considered to report for dark energy.  Literature \cite{Nojiri,rev10,rev11} also argued the reasons why modified gravity approach is extremely attractive in the appliances for late-time acceleration of the universe and to understand the problems of dark energy.
The modified gravity theories,  namely, $f (R)$ (where $R$ being the Ricci scalar curvature) \cite{Nojiri1},  $f(R,T_1)$ gravity (where  $T_1$ being the trace of the stress-energy tensor)) \cite{harko, beesham1},   Horava-Lifshitz and Gauss-Bonnet \cite{Li} theories have been projected. 

Freshly, another interesting type of modified theories is emerged as $F(T)$ gravity ($T$ is torsion scalar) \cite{Maurya, Chanchal, Coley}. Such $F(T )$ gravity theories  admit the present accelerated expansion of the Universe. The basic review on various Teleparallel gravity and the corresponding cosmological and astrophysical applications is given in the literature \cite{Rev1}. The basic works on dynamical analysis of $f(T)$ gravity are studied in the Ref. \cite{Rev2, Rev3}.  It is fascinating to reminder in $F(T)$ gravity the equations of motion are second order in contrast with GR where the field equations are fourth order equations. Noteworthy cosmological properties of modified $ F(T )$ gravity models have been studied in the literature \cite{Cai,Bamba1,Ferraro_T}. The $F(T)$-gravity  \cite{Barrow_T} theories also accumulate the accelerated expansion of the Universe and even the primordial inflationary phase. In recent times such a modification of Einstein's theory is found to describe some of the observed features relevant for cosmology and astrophysics \cite{Cappo_T}. In $F(T)$ gravity, literature \cite{baffou} takes into account an interacting cosmological fluid described by generalized Chaplygin gas with viscosity.  The stationary scenario between dark energy and dark matter is also studied \cite{Capozzielo_T} in $F(T)$ theory. Analytical solutions in the case of charged black hole can be found in $F(T)$ gravity \cite{Capozzielo_T2}. Cosmological predictions of several $f(T)$ models  are investigated in Ref. \cite{Rev4}  using current cosmological observations. Ref {\cite{Rev5} investigates  $f(T)$ cosmology in both the background and perturbation level  and present  the equivalent one-parameter family of $f(T)$ models. Cosmological perturbations in $f(T)$ gravity are investigated in Literature \cite{Rev6}  and find that $f(T)$ gravity is free of massive gravitons.  The $F(T) $ gravity and its cosmological relevance have achieved a lot of attention in the literature for early as well as late time evolution of the universe \cite{Wu,Bamba_T,Nashed}.

  The dissipative processes \cite{misner,zimdahl1} in the early universe can guide to diverge  from perfect fluid assumptions \cite{pavon,lima}, which consent the presence of  viscosity \cite{rev12,rev16} to investigate the evolution of the universe. In the early universe, viscosity \cite{rev13,rev17}  may invent due to various processes e.g.,  decoupling of matter from radiation during the recombination era, particle collisions involving gravitons and formation of galaxies \cite{barrow}. A non-negligible bulk viscous stress is also important at late-time evolution of the universe \cite{pavon1,rev15}.  Eckart \cite{eckart} first formulated a relativistic theory of viscosity. However the theory of Eckart suffers from shortcomings, namely, causality and stability \cite{hiskock}. Subsequently, Israel and Stewart \cite{israel} developed a relativistic formulation of the theory which is termed as transient or extended irreversible thermodynamics ( in short $\it{EIT}$) which provides a acceptable substitution of the Eckart theory. Using the transport equations obtained from  $\it{EIT}$, cosmological solutions are investigated in Einstein's gravity \cite{arbab,pradhan,colistete}.  Thus it is essential to investigate cosmological solutions in $F(T)$ gravity with viscosity depicted by  ${\it EIT}$. 
 
In this article, we will study $F(T) $ torsion gravity theory and the corresponding cosmological solutions  in the presence of the bulk viscosity illustrated by Eckart theory and Truncated Israel Stewart theory. We are intended to determine exact solutions and as well as stability of the solutions in $F (T )$ gravity  for flat isotropic and homogeneous space time. Different types of analytical models such as exponential expansion, power law expansion, hybrid expansion, quasi-exponential expansion, emergent universe model,   cyclic model etc are used in literature [17] to explain evolution of the universe. Exponential expansions are considered here and it is supported by cosmic observations [19]. The sketch of this paper is as follows: in sec. 2, we give the appropriate field equations in $F(T)$ theory of gravity. In sec. 3, cosmological solutions are endorsed. In sec. 4, we summarize the results achieved.

\section{Field Equations in $F(T)$ gravity theory}
In the $F(T)$ theory gravity formalism action is given by \cite{harko}
\begin{equation}
I=\frac{1}{2}\int d^4x |h| \left( T+f(T)+L_m \right),
\end{equation}
 where we choose $F(T)= T + f(T)$ and the unit $ 8\pi G=1, \;  c=1$. Here $T$ represents torsion scalar, $f(T)$ is a differentiable function of torsion,  $h=det(h^{i}_\nu)=\sqrt{-g}$ and the matter Lagrangian corresponds to $L_m$. The torsion scalar is represented by $T=S^{\mu\nu}_{\sigma}T^{\sigma}_{\mu\nu}$. In this framework the gravitational field appears corresponding to torsion tensor $(T^{\sigma}_{\mu\nu})$ defined as $T^{\sigma}_{\mu\nu} = \Gamma^{\sigma}_{\nu\mu}-\Gamma^{\sigma}_{\mu\nu} \equiv h^{\sigma}_i(\partial_{\mu}e^{i}_{\nu}-\partial_{\nu}e^{i}_{\mu})$, where the Weitzenbock connection ($\Gamma^{\sigma}_{\mu\nu}$) is given by $\Gamma^{\sigma}_{\mu\nu} \equiv h^{\sigma}_{i}\partial_{\mu}e^{i}_{\nu}$ . The contracted  scalar form of torsion tensor yields
\begin{equation}
T \equiv \frac{1}{4} T^{\sigma\mu\nu}T_{\sigma\mu\nu}+\frac{1}{2} T^{\sigma\mu\nu}T_{\mu\sigma\nu}-T^{\sigma\mu}_{\mu}T^{\nu}_{\sigma\nu} \;.
\end{equation}
The Superpotential is defined as 
\begin{equation}
S^{\mu\nu}_{\sigma} = \frac{1}{2} (K^{\mu\nu}_{\sigma}+\delta^{\mu}_{\sigma} T^{\eta \nu}_{\eta}- \delta^{\nu}_{\sigma} T^{\eta \nu}_{\nu})\;.
\end{equation}
The Superpotential is skew symmetric ($S^{\mu\nu}_{\sigma}=-S^{\nu\mu}_{\sigma}$) and the superpotential tensor plays a crucial role in the dynamics of teleparallel gravity.
The contortion tensor is defined by 
\begin{equation}
K_{\sigma\mu\nu} =  \frac{1}{2}(T_{\nu\sigma\mu}-T_{\sigma \mu \nu} - T_{\mu\sigma\nu}) \;.
\end{equation}
Varying the action, the field equations yield
\begin{equation}
h^{-1}\partial_{\mu}(h h^{\sigma}_{i} S^{\mu\nu}_{\sigma} ) [1+f_T] + h^{\sigma}_{i} S^{\mu\nu}_{\rho} \partial_{\mu}(T)f_{TT}-[1+f_T]h^{\eta}_{i}T^{\sigma}_{\mu\eta}S^{\nu\mu}_{\sigma} + \frac{1}{4} h^{\nu}_{i} [T + f(T)] = \frac{1}{2} h^{\sigma}_{i} T^{\nu}_{\sigma}
\end{equation}
where $f_T = \frac{df}{dT}$, $f_{TT}= \frac{d^2f}{dT^2}$ and $T^{\nu}_{\sigma}$ represents total energy momentum tensor.
We consider the  flat homogeneous and isotropic space-time given by Friedmann-Robertson-Walker (FRW) metric
\begin{equation}
ds^2= dt^2 - a^2 (t) \left[dr^2+ r^2 (d\theta^2+sin^2\theta d\phi^2 ) \right] ,                       
\end{equation}     
where $a(t)$ is the scale factor of the universe.  The energy-momentum tensor of the metric as given by 
\begin{equation}
T_{\mu\nu}=(\rho+ \bar{p})u_{\mu} u_{\nu} - \bar{p} g_{\mu\nu} ,
\end{equation}
where $\rho$ is energy density of the universe, $\bar{p}$ is the effective pressure, $u^{\mu}$ is the four velocity and  $u^{\mu}u_{\mu}=1$. 
Where torsion scalar $T$ is given by
\begin{equation}
T = -6 H^2,
\end{equation} 
The modified field equations in the framework of $f(T)$ gravity \cite{Surajit} are given by 
\begin{equation}
H^2 =\frac{\rho}{3} + \frac{Tf_T}{3}-\frac{f(T)}{6}, 
\end{equation}
\begin{equation}
\dot{H}(1+2Tf_{TT}+f_T) =-\frac{1}{2} (\rho + \bar{p}).
\end{equation}
To simplify the highly nonlinear field equations we require  a specific form of $f(T)$. In this paper we consider $f(T)= \gamma T + \alpha (-T)^n$,  where $\gamma$ and $\alpha$ are arbitrary constants. The field equations yield 
\begin{equation}
H^2 =\frac{\rho}{3} + \frac{1}{6}\left[\gamma T + (2n-1) \alpha (-T)^n \right], 
\end{equation}
\begin{equation}
\dot{H}\left[1+\gamma -(2n-1)\alpha n (-T)^{n-1} \right] =-\frac{1}{2} (\rho + \bar{p}).
\end{equation}
 Using Eq. (8), the expression of energy density and effective pressure in $F(T)$ $(= (1+ \gamma) T + \alpha (-T)^n$ gravity  respectively yield   
\begin{equation}
\rho = 3\left(1+\gamma\right)H^2 -\frac{1}{2} \alpha (2n-1) 6^n H^{2n} ,
\end{equation} 
\begin{equation}
 \bar{p} = -3\left(1+\gamma \right) H^2 +\frac{1}{2}\alpha (2n-1) (6 H^2)^{n} - 2\dot{H} \left[ 1+\gamma-(2n-1)\alpha n (6H^2)^{n-1}\right].     
\end{equation}                   
Let the effective pressure  ($\bar{p}$) contain two parts : $\bar{p}=p+\Pi$, where $p$ is the isotropic pressure of the universe and   $\Pi \;(\leq 0) $ is the bulk viscous pressure. Here we consider linear equation of state (EoS) of the cosmic  fluid, i.e., 
$p=\omega \rho $, where $\omega \;(1\geq\omega\geq -1) $ represents EoS parameter. The expression of the bulk viscous pressure is given by  
\begin{equation}
\Pi = -3(1+\omega)(1+\gamma) H^2 + \alpha(n-\frac{1}{2})( 1+\omega) (6H^2)^{n}  -2\dot{H}\left[ 1+\gamma - (2n-1)\alpha n (6H^2)^{n-1}\right] . 
\end{equation}
Physically acceptable cosmological solutions in the presence of bulk viscosity are permitted for $\rho>0$, $\Pi<0$ and $|\Pi|<< \rho$.   Viable exponential cosmological model in flat FRW space time  with $F(T)=(1+\gamma)T+\alpha (-T)^n$ gravity is permitted for following constraints on coupling parameters: case (i): $\gamma>-1$, $\alpha \leq 0 $ and $n\geq \frac{1}{2}$, case (ii)  $\gamma>-1$, $\alpha \geq 0 $ and $n\leq \frac{1}{2}$, case (iii)  $\gamma>-1$, $H<\left[\frac{1+\gamma}{\sqrt{6}\alpha(2n-1)}\right]^{\frac{1}{2n-2}} $ and $(n-\frac{1}{2})\alpha >0$. 
 By limiting the coupling parameters ($n\rightarrow \frac{1}{2}, \alpha \rightarrow 0$ and $\gamma \rightarrow 0$) of $f(T)$ gravity one can recover solutions of general relativity. Hence,  $f(T)$ gravity is like a tweak to the general relativity equation that allows for more flexibility in explaining cosmic evolutions $\cite{Rev1}$.
  It is worthy to note, one can recover the standard cosmological models with viscosity \cite{rev13,rev14,zimdahl} from field Eqs. (13)-(15) for (i) $\gamma=\alpha =0$, (ii) $n=1,\; \gamma=\alpha$, (iii) $n=\frac{1}{2},\; \gamma =0$. The present article we have considered a generalized form  ($F(T)=(1+\gamma)T +\alpha (-T)^n$) to build cosmological solutions, However  in  paper $\cite{Chanchal}$ where we have considered $F(T)=(1+\gamma)T +\alpha T^2$. In the following section, we shall study cosmological solutions in the presence of bulk viscosity in $F(T)$ gravity. 
\section{Cosmological  Solutions}
 The bulk viscous stress satisfies following transport equation \cite{eckart, israel}
\begin{equation}
\Pi +\tau\dot{\Pi}=-3\zeta H ,            
\end{equation}
where the parameter $\zeta (\geq 0)$ is the co-efficient of bulk viscosity, the parameter $\tau (\geq 0)$ is the relaxation time.  Eckart theory is recovered from Eq. (16) for   $\tau=0$.  
The set of Eqs. (13)-(16) are employed to obtain cosmological solutions in $F(T)$ gravity with bulk viscosity. The systems of equations are not closed.  It is known that the coefficient of bulk viscosity  and relaxation time are, in general, functions of time (or of the energy density). We, therefore, consider following relations \cite{brevik,meng,jou}
\begin{equation}
\zeta=\beta\;\rho^s ,\;\;\;\; \tau=\beta \;\rho^{s-1} ,
\end{equation}            
where   $\beta\;(>0)$ and $s\;(>0)$ are  constants. 
The conservation equation for particle number can be written as 
\begin{equation}
\dot{n_1}+3H n_1 =0,
\end{equation}
 where $n_1$ is the particle number density.  
The equilibrium variable $S$ is given by 
\begin{equation}
\dot{S}=-\frac{3 H \Pi}{n_1 T_1} ,
\end{equation}
where $S$ and $T_1$ are the specific entropy and temperature of the universe respectively. Non-equilibrium effective specific entropy is given by 
\begin{equation}
S_{eff} = S-\left(\frac{\tau}{2n_1T_1\zeta}\right)
\end{equation}
 Here we consider the temperature $(T_1)$ of the universe satisfying barotropic form as $T_1 = T_0 H^{\frac{\omega}{1+\omega}}$, where $T_0$ is a constant.   
\subsection{Eckart Theory ($\tau=0$):} Using Eqs. (13), (15) and (16) for $\tau=0$ we get
\begin{equation}
\frac{2}{3}\dot{H} (A_\gamma-n B_n  H^{2n-2}) + (A_\gamma H^2 -B_n H^{2n}) \left(1+\omega -3\beta H(A_\gamma H^2-B_n H^{2n})^{s-1}\right) =0,
\end{equation}                
where $A_\gamma=3(1+\gamma)$ and $B_n = 3\alpha(2n-1)6^{n-1}$. The evolution of the universe in Eckart theory can be obtained by using Eq. (21).
 As the above Eq. (21) is highly non-linear and the relativistic solution cannot be expressed in known general analytic form, we consider power law and exponential solution as a special case for simplicity.\\
To study power law solution ($a(t)=a_0 t^D$, where $a_0$ and $D$ are constants) in Eckart theory Eq. (21) yields,
\begin{equation}
\left[(1+\omega)D-\frac{2}{3}\right]\frac{DA}{t^2} +\frac{B_n D^{2n-1}}{t^{2n}} \left[\frac{2n}{3}-(1+\omega)D\right]-\frac{3\beta D}{t}\times\left[\frac{AD^2}{t^2}-\frac{B_n D^{2n}}{t^{2n}}\right]^s= 0  . 
\end{equation}
Using Eq. (22), we note that power law solutions are permitted in $F(T)$ gravity. We note the following cases:\\
Case (i) $\alpha=0$ or $n=\frac{1}{2}$ and $s=\frac{1}{2}$: In this case the  energy density and bulk viscosity respectively yields $\rho=\rho_0t^{-2}$ and $\Pi=-\Pi_0 t^{-2}$, where $\rho_0=3(1+\gamma) D^2 $ and $\Pi_0=(1+\gamma)D(3(1+\omega)D-2)$. The  power law exponent  $D=\frac{2(1+\gamma)}{3(1+\gamma)(1+\omega)-3\sqrt{3}\beta\sqrt{1+\gamma}}$,  leads the  lower boundary values of viscous constant $\beta> \frac{\sqrt{3(1+\gamma)}(3\omega+1)}{9}$ for accelerated universe.\\ 
Case (ii) $n=1$ and $s=\frac{1}{2}$: In this case the  energy density and bulk viscosity respectively yields $\rho=\rho_0t^{-2}$ and $\Pi=-\Pi_0 t^{-2}$, where $\rho_0=3(1+\gamma -\alpha) D^2$ and $\Pi_0=(1+\gamma-\alpha)D(3(1+\omega)D-2)$. The power law exponent $D=\frac{2(1+\gamma -\alpha)}{3(1+\gamma-\alpha)(1+\omega)-3\sqrt{3}\beta\sqrt{1+\gamma-\alpha}}$. The power law accelerated universe ($D>1$) is permitted for $ \beta > \frac{\sqrt{3(1+\gamma-\alpha)}(3\omega+1)}{9}$. \\
Case (iii) $\gamma=-1$ and $s=\frac{2n-1}{2n}$: The expression of energy density and bulk viscosity respectively yields $\rho=\rho_0 t^{-2n}$ and $\Pi=-\Pi_0 t^{-2n}$, where $\rho_0=-\frac{1}{2}\alpha (2n-1) 6^n D^{2n}$ and $\Pi_0=-\alpha D^{2n-1} 6^{n-1} (2n-3D(1+\omega))$. The power law exponent $D=\frac{2n}{3(1+\omega)-9\beta(3\alpha(1-2n)6^{n-1})^{-\frac{1}{2n}}}$. It exhibits accelerated universe for $\beta> \frac{3(1+\omega)-2n}{9 (3\alpha(1-2n)6^{n-1})^{-\frac{1}{2n}} }$. \\
In power law model ($a(t)\sim t^D$) in the expression of energy density and bulk viscous pressure yields respectively
\begin{equation}
\rho = \rho_1 t^{-2}+ \rho_2 t^{2n}, \;\;\;\;\;\ \Pi=-\Pi_1 t^{-2} -\Pi_2 t^{-2n}
\end{equation}
where $\rho_1=3(1+\gamma) D^2$, $\rho_2 = -\frac{\alpha}{2}(2n-1)6^n D^{2n}$, $\Pi_1=D(1+\gamma)(3(1+\omega)D-2)$ and $\Pi_2=(2n-1)\alpha 6^{n-1} D^{2n-1} \times (2n-3(1+\omega)D)$.  In power law model ($a(t)\sim t^D$) in Eckart theory particle density evolve as $n_1=n_0 t^{-3 D}$, where $n_0$ and $D$ are positive constants. In Eckart theory entropy evolve as 
\begin{equation}
 S_{eck} = S_0 +(S_1 + S_2 t^{2(1-n)})\times t^{3D-\frac{2+\omega}{1+\omega}}, 
\end{equation}
where $S_0>0 $ is a integration constant, $S_1=\frac{3(1+\gamma)}{n_0 T_0}\times D^{\frac{2+\omega}{1+\omega}}\times (3D(1+\omega)-2)\times \frac{1+\omega}{3D(1+\omega)-2-\omega}$, and $S_2 =\frac{3\alpha (2n-1)}{n_0} \times 6^{n-1} (2n-3D(1+\omega))\times D^{\frac{2n+(2n-1)\omega}{1+\omega}}\times \frac{1+\omega}{(3D-2n)(1+\omega)+\omega}$. In Eckart theory positive evolution of entropy  can be obtained for positive values of $S_0, S_1 $ and $S_2$. We note that $S_1>0$ for (i) $D> \frac{2}{3(1+\omega)}$ and $D> \frac{2+\omega}{3(1+\omega)}$ or (ii)  $D< \frac{2}{3(1+\omega)}$ and $D < \frac{2+\omega}{3(1+\omega)}$. We also note that $S_2>0$ for (i) $n> \frac{1}{2}$, $\alpha>0$, $D< \frac{2n}{3(1+\omega)}$ and $D> \frac{2n}{3}-\frac{\omega}{3(1+\omega)}$ or (ii)  $n< \frac{1}{2}$, $\alpha<0$, $D< \frac{2n}{3(1+\omega)}$ and $D> \frac{2n}{3}-\frac{\omega}{3(1+\omega)}$, or (iii) $n< \frac{1}{2}$, $\alpha>0$, $D> \frac{2n}{3(1+\omega)}$ and $D> \frac{2n}{3}-\frac{\omega}{3(1+\omega)}$, or (iv) $n> \frac{1}{2}$, $\alpha<0$, $D> \frac{2n}{3(1+\omega)}$ and $D> \frac{2n}{3}-\frac{\omega}{3(1+\omega)}$, or (v) $n< \frac{1}{2}$, $\alpha>0$, $D< \frac{2n}{3(1+\omega)}$ and $D< \frac{2n}{3}-\frac{\omega}{3(1+\omega)}$, or (vi) $n> \frac{1}{2}$, $\alpha<0$, $D< \frac{2n}{3(1+\omega)}$ and $D< \frac{2n}{3}-\frac{\omega}{3(1+\omega)}$, or (vii) $n> \frac{1}{2}$, $\alpha<0$, $D> \frac{2n}{3(1+\omega)}$ and $D> \frac{2n}{3}-\frac{\omega}{3(1+\omega)}$, or (viii) $n< \frac{1}{2}$, $\alpha>0$, $D> \frac{2n}{3(1+\omega)}$ and $D> \frac{2n}{3}-\frac{\omega}{3(1+\omega)}$.\\
The left panel of Fig. (1) shows power law exponent $D$ versus EoS parameter $\omega$ for a given set of other parameter. The white regions in the plot are suitable for positive entropy creation in Eckart theory. In the white region all the three coefficients ($S_1, S_2, S_3$) are positive. The left panel shows that in $F(T)$ gravity power law accelerated ($D>1$ or $q<0$) evolution are suitable for higher values of EoS parameter $\omega$.
\begin{figure}[ph]
\centerline{\includegraphics[width=1.7in]{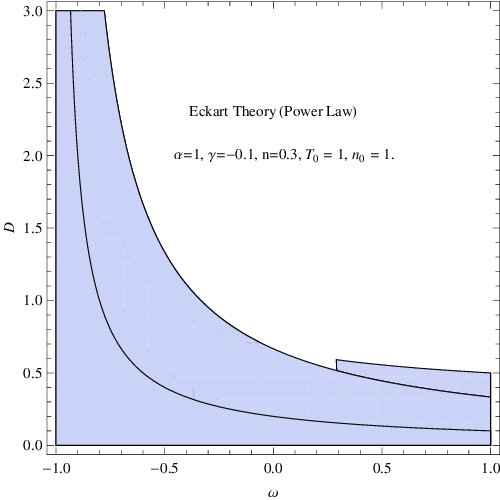}\includegraphics[width=1.7in]{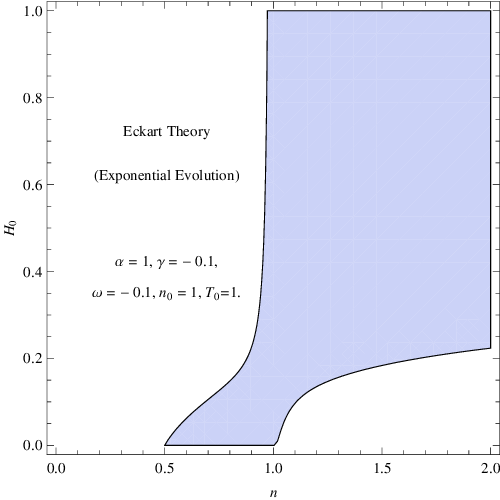}\includegraphics[width=1.7in]{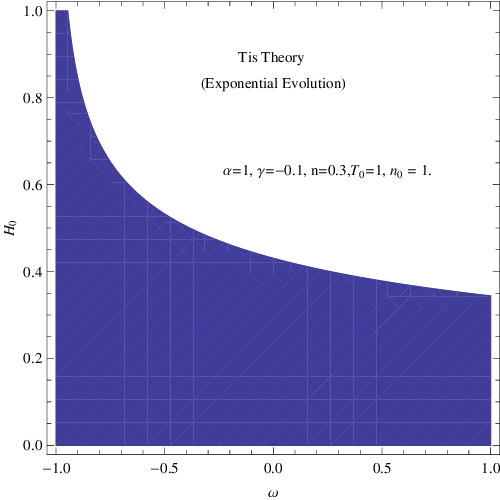}}
\vspace*{8pt}
\caption{ In the region plots shadow regions are unsuitable for model building.  The left panel shows suitable ranges of EoS parameter $\omega$ and power law exponent $D$ for positive entropy evolution in Eckart theory for a given set of other parameters.  The middle panel shows suitable ranges of exponential parameter $H_0$ and exponent $n$ in $F(T) $
$(=(1+\gamma)T+\alpha(-T)^n)$)  gravity  for positive entropy evolution in Eckart theory for a given set of other parameters.The right panel shows suitable ranges of  EoS parameter $\omega$ and exponential parameter $H_0$  for positive entropy evolution in TIS theory for a given set of other parameters.  \protect\label{fig1}}
\end{figure}
\\
 The exponential  expansion ($a(t) = a_0 \exp[H_0 t]$,  where $a_0$ and $H_0$ are positive constants) of the universe with viscosity  can be obtained  in $F(T)$ gravity by setting $H=H_0=const.$  In exponential model the expression of energy density and bulk viscous stress yields respectively 
\begin{equation}
\rho = 3\left(1+\gamma\right)H_0^2 -\frac{1}{2} \alpha (2n-1) 6^n H_0^{2n} ,
\end{equation}
\begin{equation}
\Pi = -3(1+\omega)(1+\gamma) H_0^2 + \alpha(n-\frac{1}{2})( 1+\omega) (6H_0^2)^{n}  . 
\end{equation}
In this case the equation (21) yields 
\begin{equation}
 (A_\gamma H_0^2 -B_n H_0^{2n}) \left(1+\omega -3\beta H_0(A_\gamma H_0^2-B_n H_0^{2n})^{s-1}\right) =0
\end{equation}
 The exponential acceleration of the universe in $F(T)$ gravity is permitted  and the following special cases are noted in Table 1. To study stability analysis Eq. (21) can be rewritten as
\begin{equation}
\dot{H} =f(H)=\frac{3 (A_\gamma H^2 -B_n H^{2n}) \left(3\beta H(A_\gamma H^2-B_n H^{2n})^{s-1}-1-\omega\right) }{ 2 (A_\gamma-n B_n  H^{2n-2}) }, 
\end{equation}
where $A_\gamma \neq n B_n  H^{2n-2}$. In Eckart theory fixed or equilibrium point \cite{Strogatz}  can be obtained for $\dot{H}=0$ or $H=H_0=const.$ which yields $3\beta H_0(A_\gamma H_0^2-B_n H_0^{2n})^{s-1}=1+\omega $. The fixed points represent de Sitter type evolution of the universe even in the presence of matter. The stability of the fixed points are obtained from the expression $f'(H_0) =\frac{df(H)}{dH}|_{H=H_0}$, i.e., 
\begin{equation}
f'(H_0)=\frac{9\beta (A_\gamma H_0^2 -B_n H_0^{2n})^s}{2(A_\gamma H_0^2-n B_n H_0^{2n})} H_0^2 +3(s-1) H_0 (1+\omega), 
\end{equation}
The fixed points ($\dot{H}=0$)  represent equilibrium or steady exponential solution. Equilibrium or fixed points is defined to be stable if all sufficient small deviation away from it damp out in time. Conversely in unstable equilibrium, all sufficient small deviation grows in time. The fixed points ($H_0$) are stable for  $f'(H_0)<0$ and that are unstable for $f'(H_0)>0$. Stability analysis in Eckart theory for accelerated exponential expansion are summarized in Table 1.
\begin{table}[ht]
\tbl{ Stability of the exponentially accelerated $(H_0 >0)$ expansions in Eckart theory}
{\begin{tabular}{@{}cccc@{}} \toprule
$Cases$ & Expression of Fixed points ($H_0$)& Stability & Conditions for Stability\\ \colrule
(i) $s=0,$ & $ \frac{\beta\pm \sqrt{\beta^2-2\alpha(1+\omega)^2(1+\gamma)/3}}{2(1+\omega)(1+\gamma)}$& Stable &  $\beta>(1+\omega)\sqrt{2\alpha(1+\gamma)/3},\;\gamma>-1$,\\
$n=0$. &&& $H_0>\frac{\beta}{2(1+\gamma)(1+\omega)},\;\beta>0,\;\omega>-1.$\\
(ii) $s=0,$ & $\frac{\beta}{(1+\omega)(1+\gamma)}$& Stable & $\beta>0,\;\gamma>-1,\; \omega>-1.$   \\
$n=\frac{1}{2}$. &&& \\\\
(iii) $s=0,$ & $ \frac{\beta}{(1+\omega)(1+\gamma-\alpha)} $& Stable & $\beta > 0$, $\alpha<1+\gamma,$   \\
$n=1$. &&& $\omega>-1$. \\
(iv) $s=0,$ & $ \frac{(1+\gamma)\pm \sqrt{(1+\gamma)^2 -8\sqrt{6}\alpha\beta/(1+\omega)}}{4\sqrt{6}\alpha} $& Stable & $0<\beta< \frac{(1+\omega)(1+\gamma)^2}{8\sqrt{6}\alpha},$  \\
$n=\frac{3}{2}$. &&& $\omega>-1,\; \alpha>0$. \\
(v) $s=1 $ & $ \frac{1+\omega}{3\beta} $ & Unstable & $\beta>0,\;\omega>-1.$   \\
 (vi) $s=2,$ & $ \left[\frac{(1+\omega)}{9\beta (1+\gamma)} \right]^{\frac{1}{3}}$& Unstable  &  $\beta >0, \; \gamma>-1.$\\
$n=\frac{1}{2}$. &&& $\omega>-1$. \\ 
 (vii) $s=2,$ & $ [\frac{1+\omega}{9\beta (1+\gamma -\alpha) }]^{\frac{1}{3}}$& Unstable  &  $\beta >0,\; \alpha<1+\gamma.$ \\
$n=1$. &&& $\omega>-1$. \\ 
(viii) $\alpha=0,$  & $ \left[\frac{1+\omega}{3^s \beta (1+\gamma)^{s-1} }\right]^{\frac{1}{2s-1}}$ & Stable &  $\beta >0,\gamma>-1.$ \\
or $n=\frac{1}{2}$. &&& $s<\frac{1}{2},\;\omega>-1$. \\ 
(ix) $\gamma=-1$. & $ \left[\frac{1+\omega}{3^s \beta (\alpha(1-2n)6^{n-1})^{(s-1)} }\right]^{\frac{1}{2n(s-1)+1}}$ & Stable &  $\omega>-1,\;\beta >0, \;n>\frac{1}{2},\;\alpha < 0.$ \\
  &&& or $\omega>-1,\;\beta >0, \;n<\frac{1}{2},\;\alpha > 0.$ \\ \botrule
\end{tabular}}
\end{table}
 One can also study stability analysis by plotting $H_0$ and $f'(H_0)$ for a given values of other parameters  in Eckart theory . Using Eqs. (28)-(29), the directional plots $H_0$ vs $f'(H_0)$  are shown in Figure (2) for a given set of other parameters. The downward arrows in the plots lead to the stable accelerated exponential evolution whereas upward arrows lead to the unstable evolution and the regions without arrows are unsuitable for accelerated expansion. The left panel shows stability analysis in Eckart theory for $s=0,\;n=0 $ with a given set of other parameters. For $s=0,\;n=0 $, stable exponential acceleration is suitable for lower values of EoS state parameter $\omega$ and higher values of bulk viscous constant $\beta$  in $F(R)$ gravity. The middle panel shows stability analysis in Eckart theory for $\alpha=0,$ or $\;n=\frac{1}{2} $ with a given set of other parameters. For $\alpha=0,\;n=\frac{1}{2} $, stable exponential acceleration is suitable for lower values of EoS state parameter $\omega$ and lower values of bulk viscous exponent $s$  in $F(R)=(1+\gamma) T$ gravity. The right panel shows the stability analysis of  exponential evolution in $F(T)(=\alpha(-T)^n)$ gravity $(\gamma =-1)$  for different values of $\alpha$ and $n$ for a given set of other parameters. For $\gamma=-1 $, stable exponential accelerated  is suitable for (i) $\alpha>0,\;n<\frac{1}{2}$ (ii) $\alpha<0,\;n>\frac{1}{2}$  in $F(T)(=\alpha(-T)^n)$.
\begin{figure}[ph]
\centerline{\includegraphics[width=1.7in]{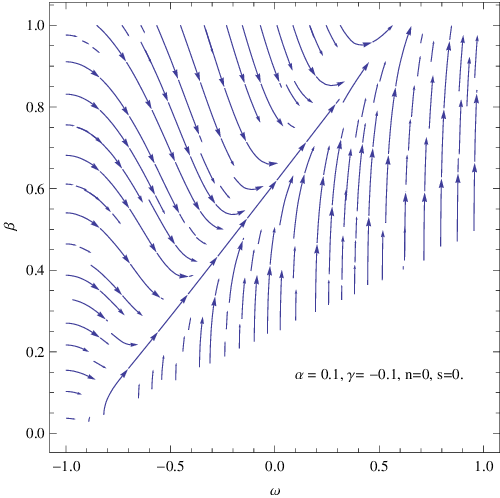}\includegraphics[width=1.7in]{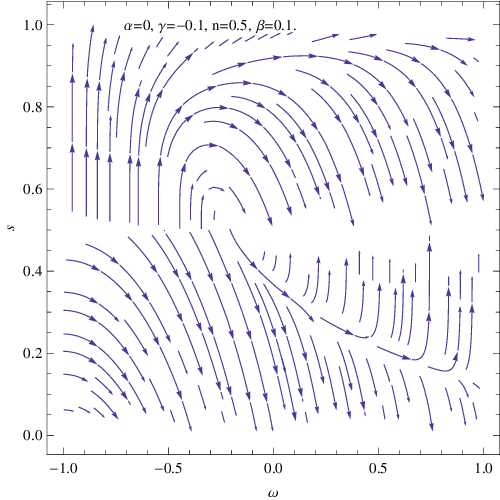}\includegraphics[width=1.7in]{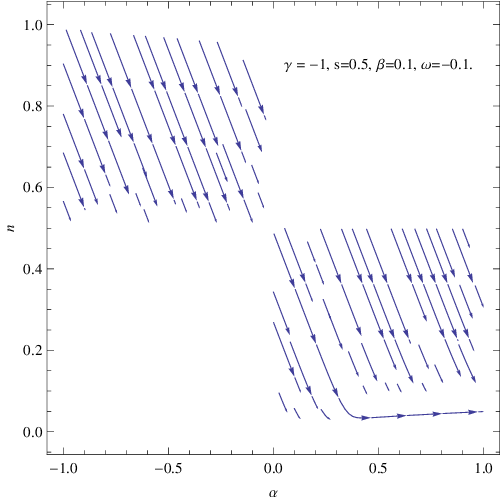}}
\vspace*{8pt}
\caption{ The left panel shows stability analysis in Eckart theory for $s=0,\;n=0 $ with a given set of other parameters. The middle panel shows stability analysis in Eckart theory for $\alpha=0,$ or $\;n=\frac{1}{2} $ with a given set of other parameters. The right panel shows the stability analysis of  exponential evolution in $F(T)(=\alpha(-T)^n)$ gravity $(\gamma =-1)$  for different values of $\alpha$ and $n$ for a given set of other parameters.  \protect\label{fig2}}
\end{figure}
In exponential expansion in Eckart theory particle density evolve as $n_1=n_0 e^{-3 H_0 t}$, where $n_0$ and $H_0$ are positive constants. In Eckart theory specific entropy evolve as 
\begin{equation}
 S_{eck} = S_0 + \frac{(1+\omega)}{n_0 T_0}\times H_0^{-\frac{\omega}{1+\omega}} \times [3(1+\gamma)H_0^2-\frac{\alpha}{2}(2n-1)6^n H_0^{2n}]\times e^{3H_0 t}, 
\end{equation}
where $S_0 $ is a integration constant. In Eckart theory positive evolution of entropy  is obtained either for  (i) $n\neq \frac{1}{2}$, $\alpha \neq 0 $, $0<H_0< \frac{1}{\sqrt{6}} \times [\frac{1+\gamma}{\alpha (2n-1)}]^{\frac{1}{2n-2}}$, $\gamma>-1$ and $\omega > -1$ or (ii) $n= \frac{1}{2}$ or $\alpha = 0 $ , $H_0 >0$, $\gamma>-1$ and $\omega > -1$. 
The values of positive entropy generation is confirmed for positive values of the coefficient of $e^{3H_0t}$ in Eq. (30). The middle panel of Fig. (1) shows exponential constant $H_0$ versus exponent of $F(T)$ gravity $n$ for a given set of other parameter. The white regions in the plot are suitable for positive entropy creation in Eckart theory. The middle panel shows suitable ranges of exponential parameter $H_0$ and exponent $n$ in $F(T) $ $(=(1+\gamma)T+\alpha(-T)^n)$)  gravity for positive entropy evolution in Eckart theory for a given set of other parameters. It leads that (i) higher values of $H_0$ and  lower values of $n$ and (ii) lower values of $H_0$ and higher values $n$ are suitable for exponential evolution in Eckart theory.
\subsection{ Truncated Israel Stewart Theory $(\epsilon =0)$ :}
 Using equations (13)-(17) for Truncated Israel Stewart (TIS) theory  ($\epsilon = 0$) in a flat universe, we obtain
\[
\ddot{H} -\dot{H}^2\left[\frac{\alpha n (n-1)(2n-1)6^n H^{2-3}}{3(1+\gamma-(2n-1)\alpha n (6H^2)^{n-1})}\right] 
\]
\begin{equation}
+\left[\frac{\rho^{1-s}}{\beta}+ 3(1+\omega)H \right]\dot{H} + \frac{(1+\omega)\rho^{2-s}\beta^{-1}-3H\rho}{2[1+\gamma-\alpha n (2n-1)(6H^2)^{n-1}]} =0 .
\end{equation}  
   The evolution of the universe in TIS theory for linear EoS in $F(T)$ gravity can be obtained by using above Eq. (31) which is highly nonlinear to obtain a general analytic  solution of known form. We consider power law and exponential solutions for simplicity.\\
To study power law solution ($a(t)=a_0 t^D$, where $a_0$ and $D$ are constants) in TIS theory Eq. (31) yields,
\[
2D-3(1+\omega)D^2 -\frac{t\;D}{\beta}\left[\rho_3 t^{-2} +\rho_4 t^{-2n} \right]^{1-s}-\frac{2\alpha n (n-1)(2n-1)6^{n-1}D^{2n-1}t^{2-2n}}{1+\gamma-(2n-1)\alpha n 6^{n-1}D^{2n-2}t^{2-2n}} + \]
\begin{equation}
\frac{\beta^{-1} (1+\omega)  (\rho_3 t^{-2}+\rho_4 t^{-2n})^{2-s} -3Dt^{-1}(\rho_3 t^{-2} + \rho_4 t^{-2n})}{2(1+\gamma-(2n-1)\alpha n 6^{n-1} D^{2n-2} t^{2-2n})}\times t^3= 0,  
\end{equation}
where $\rho_3=\rho_1 D^2=3(1+\gamma)$ and $\rho_4=\rho_2 D^2=-3\alpha (2n-1) 6^{n-1}$. Using the Eq. (32), we note the power law solutions in the following  cases:\\
Case (i) $\alpha=0$ or $n=\frac{1}{2}$ and $s=\frac{1}{2}$: The  energy density and bulk viscosity respectively yield $\rho=\rho_1 t^{-2}$ and $\Pi=-\Pi_1 t^{-2}$, where $\rho_1=3(1+\gamma)D^2$ and $\Pi_1=(1+\gamma)D(3(1+\omega)D-2)$. The  power law exponent becomes
 $D=\frac{b_1\pm\sqrt{b_1^2-4a_1c_1}}{2a_1}$, where $a_1=\frac{3}{2}(\frac{\sqrt{3(1+\gamma)}(1+\omega)}{\beta}-3)$, $b_1=3(1+\omega)+\frac{\sqrt{3(1+\gamma)}}{\beta}$  and $ c_1=2$. Which leads to the accelerated universe for $ D> 1$.\\ 
Case (ii) $n=1$ and $s=\frac{1}{2}$: The energy density and bulk viscosity respectively yield $\rho=\rho_ 1t^{-2}$ and $\Pi=-\Pi_1 t^{-2}$, where $\rho_1=3(1+\gamma -\alpha)D^2$ and $\Pi_1=(1+\gamma-\alpha)D(3(1+\omega)D-2)$. The  power law exponent becomes
 $D=\frac{b_2\pm\sqrt{b_2^2-4a_2c_2}}{2a_2}$, where $a_2=\frac{3}{2}(\frac{\sqrt{3(1+\gamma-\alpha)}(1+\omega)}{\beta}-3)$, $b_2=3(1+\omega)+\frac{\sqrt{3(1+\gamma-\alpha)}}{\beta}$  and $ c_2=2$. Which executes the power-law accelerated universe for $D>1$. \\
Case (iii) $\gamma = -1$ and $s=\frac{2n-1}{2n}$: In this case, the  energy density and bulk viscosity respectively yield $\rho=\rho_0 t^{-2n}$ and $\Pi=-\Pi_0 t^{-2n}$, where $\rho_0=\rho_{02} D^2=-\frac{1}{2}\alpha (2n-1) 6^n D^{2n}$ and $\Pi_0=-\alpha D^{2n-1} 6^n (2n-1)$. The power law exponent yields The  power law exponent becomes
 $D=\frac{b_3\pm\sqrt{b_3^2-4a_3c_3}}{2a_3}$, where $a_3=\frac{3}{n}(3-(1+\omega)(-\alpha(2n-1)/2)^{\frac{1}{2n}}\sqrt{6})$, $b_3=-3(1+\omega)$  and $ c_3=2+\frac{n-1}{3}-\frac{1}{\beta} (-\alpha(2n-1)/2)^{\frac{1}{2n}}\sqrt{6}$. Energy density becomes positive $(\rho_2>0)$ for (i) $\alpha<0$ and $n>\frac{1}{2}$ (ii) $\alpha>0$ and $n<\frac{1}{2}$.  \\
 In power law model ($a(t)\sim t^D$) in TIS theory non equilibrium specific entropy evolve as 
\begin{equation}
 S_{tis} =   S_0 +(S_1 + S_2 t^{2(1-n)})\times t^{3D-\frac{2+\omega}{1+\omega}} -\frac{D^{\frac{-\omega}{1+\omega}} }{2n_0 T_0}\times \frac{t^{3D+\frac{\omega}{1+\omega}}}{(\rho_3 t^{-2}+\rho_4 t^{-2n})}.
\end{equation}
Where $S_0>0 $ is a integration constant, $S_1=\frac{3(1+\gamma)}{n_0 T_0}\times D^{\frac{2+\omega}{1+\omega}}\times (3D(1+\omega)-2)\times \frac{1+\omega}{3D(1+\omega)-2-\omega}$, and $S_2 =\frac{3\alpha (2n-1)}{n_0} \times 6^{n-1} (2n-3D(1+\omega))\times D^{\frac{2n+(2n-1)\omega}{1+\omega}}\times \frac{1+\omega}{(3D-2n)(1+\omega)+\omega}$. In TIS theory positive evolution of entropy  leads to a viable power law expansion. Eq. (33) shows the evolution of entropy in TIS theory for power-law expansion in $F(T)$ gravity.

The exponential  expansion $(a(t) = a_0 \exp[H_0 t]$, where $a_0$ and $H_0$ are positive constants ) of the universe with viscosity  can be obtained  in $F(T)$ gravity by setting $H=H_0=const.$ In exponential model the expression of energy density and bulk viscous stress yields respectively 
\begin{equation}
\rho = 3\left(1+\gamma\right)H_0^2 -\frac{1}{2} \alpha (2n-1) 6^n H_0^{2n} ,
\end{equation}
\begin{equation}
\Pi = -3(1+\omega)(1+\gamma) H_0^2 + \alpha(n-\frac{1}{2})( 1+\omega) (6H_0^2)^{n}  . 
\end{equation}
  For exponential evolution in $F(T)$ gravity, the energy density ($\rho = const.$) and bulk viscous stress ($\Pi= const.)$  remain constants. For the exponential evolution  Eq. (31) yields 
\begin{equation}
3\beta H_0(A_\gamma H_0^2-B_n H_0^{2n})^{s-1} =1+\omega
\end{equation}
 The Eq. (36) is used to determine fixed points and it executes exponential accelerations in $F(T)$ gravity for TIS theory. We note  exponential expansion in TIS theory and results are summarized in Table 2.
  The exponential evolution admits  accelerating phase for $H_0>0$.  In the TIS theory the cosmological evolution is directed by differential equation which is second order in nature. To study stability of cosmic evolution due to equilibrium points or fixed points in $F(T)$ gravity in TIS theory, one can rewrite Eq. (31) in term of two autonomous first order differential equations which are 
\[ \dot{H}=y ,\]
\begin{equation}
\dot{y}=P(y,H)=-[3(1+\omega)H+\frac{\rho^{1-s}}{\beta}]y + \frac{3H\rho-(1+\omega)\rho^{2-s}\beta^{-1}}{2[1+\gamma-\alpha n (2n-1)(6H^2)^{n-1}]}.  
\end{equation} 
The fixed points corresponding to de-Sitter type expansion are $y=\dot{y}=0$ i.e., $H=H_0=const.$ which yields $(1+\omega)\rho^{1-s}=3\beta H_0.$  Using linearization \cite{Jordan} technique  around the fixed points the Jacobian matrix yields 
 \[
\left[
\begin{array}{cc}
0&1 \\
 \frac{3\rho}{2[1+\gamma-\alpha n (2n-1)(6H_0^2)^{n-1}]}+9H_0^2(s-1) \;\;&\;\;-3H_0[(1+\omega)+\frac{1}{1+\omega}] \\
\end{array}
\right]. \]
 Hence calculating trace $(\tau_1)$ and determinant $(\Delta)$, the stability of the fixed points can be analyzed.            
The  equilibrium points associate to the de Sitter type  evolution  in $F(T)$ gravity  with TIS theory is characterized in Table 2. 

\begin{table}[ht]
\tbl{ Stability of the exponentially accelerated $(H_0 >0)$ expansions in TIS theory for $\omega>-1,\; \gamma>-1$ and $\beta>0$.}
{\begin{tabular}{@{}cccc@{}} \toprule
$Cases$ & Expression of Fixed points ($H_0$)& Stability & Conditions for Stability\\ \colrule
(i) $s=0,$ & $ \frac{\beta\pm \sqrt{\beta^2-2\alpha(1+\omega)^2(1+\gamma)/3}}{2(1+\omega)(1+\gamma)}$& Stable Attractor &  $H_0>\frac{3\omega+4}{12\alpha \beta}$.\\
$n=0$. && Unstable Saddle& $0<H_0<\frac{3\omega+4}{12\alpha\beta}$.\\
(ii) $s=0,$ & $\frac{\beta}{(1+\omega)(1+\gamma)}$& Stable Node & $H_0>0$.    \\
$n=\frac{1}{2}$. &&& \\
(iii) $s=0,$ & $ \frac{\beta}{(1+\omega)(1+\gamma-\alpha)} $& Stable Node & $H_0>0,\; 1+\gamma>\alpha$.   \\
$n=1$. &&&  \\
(iv) $s=0,$ & $ \frac{(1+\gamma)\pm \sqrt{(1+\gamma)^2 -8\sqrt{6}\alpha\beta/(1+\omega)}}{4\sqrt{6}\alpha} $& Unstable Saddle& $\frac{1+\gamma}{3\alpha\sqrt{6}}> H_0 > \frac{1+\gamma}{4\alpha\sqrt{6}}$.  \\
$n=\frac{3}{2}$. && Stable Attractor & $H_0 > \frac{1+\gamma}{3\alpha\sqrt{6}}$ or $ H_0 < \frac{1+\gamma}{4\alpha\sqrt{6}}$.  \\
(v) $s=1 $ & $ \frac{1+\omega}{3\beta} $ & Unstable Saddle & $H_0<\frac{1}{\sqrt{6}}(\frac{1+\gamma}{\alpha(2n-1)})^{\frac{1}{2(n-1)}}, 0<n<1.$   \\
&& Unstable Saddle & $H_0<\frac{1}{\sqrt{6}}(\frac{1+\gamma}{\alpha n(2n-1)})^{\frac{1}{2(n-1)}}, n>1.$   \\
&& Stable Attractor & $H_0>\frac{1}{\sqrt{6}}(\frac{1+\gamma}{\alpha n(2n-1)})^{\frac{1}{2(n-1)}}, n>1.$   \\
&& Stable Attractor & $H_0>\frac{1}{\sqrt{6}}(\frac{1+\gamma}{\alpha (2n-1)})^{\frac{1}{2(n-1)}}, 0<n<1.$   \\
 (vi) $s=2,$ & $ \left[\frac{(1+\omega)}{9\beta (1+\gamma)} \right]^{\frac{1}{3}}$& Unstable Saddle  &  $H_0>0.$\\
$n=\frac{1}{2}$. &&&  \\ 
 (vii) $s=2,$ & $ [\frac{1+\omega}{9\beta (1+\gamma -\alpha) }]^{\frac{1}{3}}$& Unstable Saddle &  $H_0 >0,\; \alpha<1+\gamma.$ \\
$n=1$. &&& $\omega>-1$. \\ 
(viii) $\alpha=0,$  & $ \left[\frac{1+\omega}{9 \beta (1+\gamma) }\right]^{\frac{1}{2s-1}}$ & Stable Attractor&  $s<\frac{1}{2}$. \\
or $n=\frac{1}{2}$. &&Unstable Saddle& $s>\frac{1}{2}$. \\ 
(ix) $\gamma=-1$. & $ \left[\frac{1+\omega}{\beta \alpha(1-2n)6^{n} }\right]^{\frac{1}{2n(s-1)+1}}$ & Stable Attractor &  $s<\frac{2n-1}{2n}.$ \\
  &&Unstable Saddle& or $s>\frac{2n-1}{2n}.$ \\ \botrule
\end{tabular}}
\end{table}
One can also study stability analysis  in TIS theory by noting $H_0$ and $\Delta$ for a given values of other parameters. Using Eqs. (36)-(37), the directional plots are drawn between $H_0$ and $\frac{\partial P(y,H)}{\partial H}|_{H_0}$  as shown in Fig. (3). 
\begin{figure}[ph]
\centerline{\includegraphics[width=1.7in]{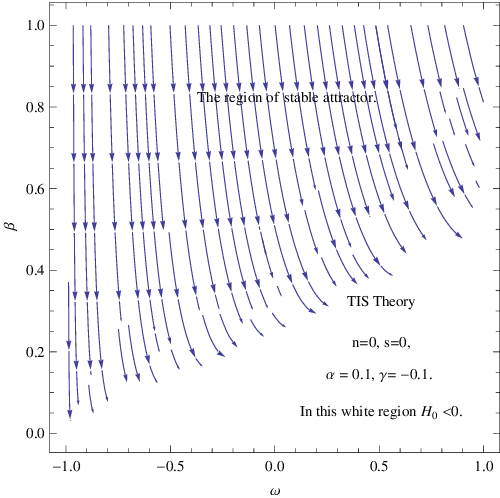}\includegraphics[width=1.7in]{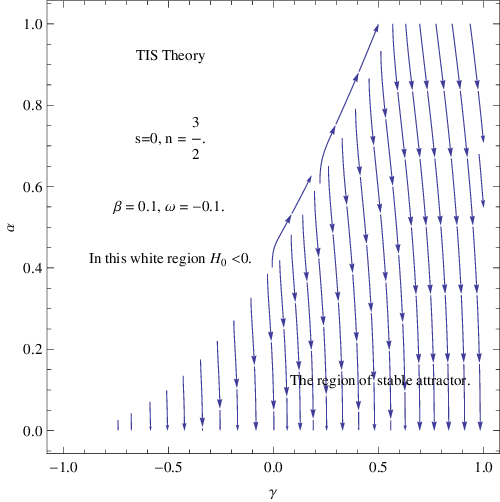}\includegraphics[width=1.7in]{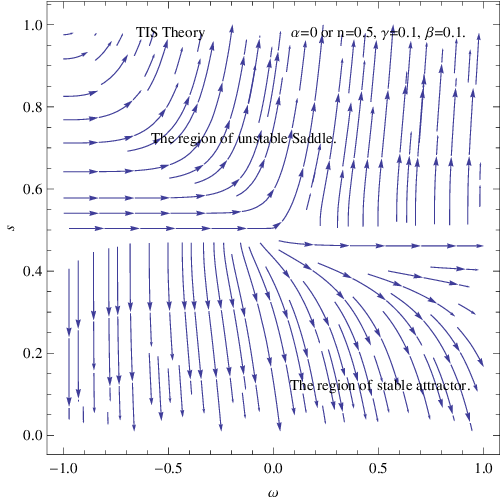}}
\vspace*{8pt}
\caption{ The left panel shows stability analysis in TIS theory for $s=0,\;n=0 $ with a given set of other parameters. The middle panel shows stability analysis in TIS theory for $s=0,$  $\;n=\frac{3}{2} $ with a given set of other parameters. The right panel shows the stability analysis in $F(T)(=(1+\gamma)T)$ gravity $(\alpha = 0)$  for different values of $s$ and $\omega$ for a given set of other parameters.  \protect\label{fig3}}
\end{figure}
 The downward arrows in the plots lead stable accelerated exponential evolution whereas upward arrows lead unstable evolution and the regions without arrows are unsuitable for accelerated expansion. The left panel shows stability analysis in TIS theory for $s=0,\;n=0 $ with a given set of other parameters. For $s=0,\;n=0 $, stable exponential acceleration is suitable for lower values of EoS state parameter $\omega$ and higher values of bulk viscous constant $\beta$  in $F(T)$ gravity. The middle panel shows stability analysis in TIS theory for $s=0,$ and $\;n=\frac{3}{2} $ with a given set of other parameters. For $s=0,\;n=\frac{3}{2} $, stable exponential acceleration is suitable for lower values of $\alpha$ and higher values of  $\gamma$  in $F(T)$ gravity. The right panel shows the stability analysis of  exponential evolution in $F(T)(=(1+\gamma)T)$ gravity $(\alpha =0)$  for different values of $s$ and $\omega$ for a given set of other parameters. For $\alpha=0 $, stable exponential accelerated  is suitable for $s<\frac{1}{2}$ in $F(T)$ gravity.

In exponential expansion in TIS theory  entropy evolve as 
\[
 S_{tis} = S_0 + \frac{H_0^{\frac{-\omega}{1+\omega}}}{n_0 T_0} \times [(1+\omega)(3(1+\gamma)H_0^2-\frac{\alpha}{2}(2n-1)6^n H_0^{2n}) \]
\begin{equation}
-\frac{1}{6(1+\gamma)H_0^2-\alpha(2n-1)6^n H_0^{2n}}]\times e^{3H_0 t} , 
\end{equation}
where $S_0 $ is a positive constant. The values of positive entropy generation is confirmed for positive values of the coefficient of $e^{3H_0t}$ in Eq. (38). The positive evolution of entropy  in TIS theory is obtained either for  (i) $n\neq \frac{1}{2}$, $\alpha \neq 0 $, $\rho> (\frac{1}{2(1+\omega)})^{\frac{1}{2}}$ or (ii) $n= \frac{1}{2}$ or $\alpha = 0 $ , $H_0 > (\frac{1}{18(1+\omega)(1+\gamma)^2})^{\frac{1}{4}}$, $\gamma>-1$ and $\omega > -1$.  However, entropy evolution in TIS theory can also be written in terms of Eckart theory as
\begin{equation}
 S_{tis} = \frac{S_0}{2\rho^2(1+\omega)} + S_{eck}(1-\frac{1}{2(1+\omega)\rho^2}).
\end{equation}
The positive evolution of entropy are suitable for exponential expansion.  The right panel of Fig. (1) shows exponential constant $H_0$ versus  $\omega$ for a given set of other parameter in $F(T)$ gravity. The white regions in the plot are suitable for positive entropy generation  in TIS theory. The right panel shows suitable ranges of exponential parameter $H_0$ and $\omega$  in $F(T) $   gravity  for positive entropy evolution in TIS theory. It leads that lower values of $H_0$ ($ > 0$) and lower  values of  $\omega$ ($<0$) are suitable for exponential evolution in TIS theory. 
 \section{Conclusion}
In this paper, we have studied   bulk viscous universe  model in $F(T)$ $(=(1+\gamma)T+\alpha (-T)^n)$ gravity in a flat FRW space time. Bulk viscosity described by  Eckart, Truncated Israel Stewart (TIS)  theories are considered  here to study geometrical and physical features of the universe.  To find cosmological solutions,  special cases are considered for  non-linearity of field equations. We have studied power law ($a(t)\sim t^D$) solutions and exponential ($a(t)\sim e^{H_0 t} $) solutions in $F(T)$ gravity. In Eckart and TIS theory, power law accelerated ($D>1$) universe are permitted in the presence of viscosity. In Eckart theory, we note the following cases for power law expansion: Case (i) $\alpha=0$ or $n=\frac{1}{2}$ and $s=\frac{1}{2}$: The  accelerated universe is permitted for a lower boundary of viscous constant as $\beta> \frac{\sqrt{3(1+\gamma)}(3\omega+1)}{9}$. Case (ii) $n=1$ and $s=\frac{1}{2}$: The power law type accelerated universe is permitted for $ \beta > \frac{\sqrt{3(1+\gamma-\alpha)}(3\omega+1)}{9}$. Case (iii) $\gamma=-1$ and $s=\frac{2n-1}{2n}$: The accelerated universe is exhibited  for $\beta> \frac{3(1+\omega)-2n}{9 (3\alpha(1-2n)6^{n-1})^{-\frac{1}{2n}} }$. We have also discussed entropy evolution and the conditions for positive entropy as shown in Fig. 1.  The left panel of Fig. (1) shows that in $F(T)$ gravity power law accelerated ($D>1$ or $q<0$) evolution are suitable for higher values of EoS parameter $\omega$ in Eckart theory. In TIS theory the expression of power law exponent and the conditions for viable power law type accelerated evolution are determined for the following cases: Case (i) $\alpha=0$ or $n=\frac{1}{2}$ and $s=\frac{1}{2}$, Case (ii) $n=1$ and $s=\frac{1}{2}$, Case (iii) $\gamma = -1$ and $s=\frac{2n-1}{2n}$. 
 The exponential expansion $(a(t)\sim e^{H_0 t})$ and the stability conditions of the solutions are studied in Eckart and TIS theory as given in Table 1 and Table 2 respectively. We have studied the stability of the exponential expansion by directional plot as shown in Figure 2 and 3 in Eckart and TIS theories respectively. In the directional plot, downwards arrows in the figures show stable exponential expansion. We note the following cases in the Fig. 2 of Eckart theory: Case (i) $s=0,\;n=0 $: The left panel shows for  stable exponential acceleration is suitable for lower values of EoS state parameter $\omega$ and higher values of bulk viscous constant $\beta$  in $F(T)$ gravity. Case (ii) $\alpha=0,$ or $\;n=\frac{1}{2} $: The middle panel shows stable exponential acceleration is suitable for lower values of EoS state parameter $\omega$ and lower values of bulk viscous exponent $s$  in $F(T)=(1+\gamma) T$ gravity. Case (iii) $(\gamma =-1)$:  The right panel shows stable exponential accelerated  is suitable for (i) $\alpha>0,\;n<\frac{1}{2}$ (ii) $\alpha<0,\;n>\frac{1}{2}$  in $F(T)(=\alpha(-T)^n)$ gravity. 
 We note the following cases in the Fig. 3 of TIS theory: Case (i) $s=0,\;n=0 $: stable exponential acceleration is suitable for lower values of EoS state parameter $\omega$ and higher values of bulk viscous constant $\beta$  in $F(T)$ gravity. Case (ii) $s=0,$ or $\;n=\frac{3}{2}$: The middle panel shows the stable exponential acceleration is suitable for lower values of $\alpha$ and higher values of  $\gamma$  in $F(T)$ gravity. Case (iii) $(\alpha =0)$ : The right panel shows the stable exponential accelerated  is suitable for $s<\frac{1}{2}$ gravity for a  in $F(T)(=(1+\gamma)T)$ gravity. 
 The middle panel of Fig. (1) in Eckart theory shows that in $F(T)$ gravity exponential expansion  accelerated ($q<0$) evolution are suitable  either for (i) higher values of $H_0$ and lower values of $n<\frac{1}{2}$ or  (ii) lower values of $H_0$ and higher values of $n>1$.
The right panel of Fig. (1) leads that higher values of $H_0$ ($ > 0$) and higher  values of  $\omega$ are suitable for exponential evolution in TIS theory. 

In conclusion, it has been observed that higher  values of  $\beta$ are suitable for both in power law type and exponential type accelerated expansion in the Eckart and TIS theory for $F(T)=(1+\gamma)T+\alpha (-T)^n)$ gravity. In $F(T)$ $(=(1+\gamma)T+\alpha (-T)^n)$  gravity power law accelerated ($D>1$ or $q<0$) evolution are suitable for higher values of EoS parameter $\omega$.  The stable exponential accelerated  is suitable for $s<\frac{1}{2}$ in $F(T)(=(1+\gamma)T)$ gravity. 
\\
\section*{Acknowledgement}
Author would like to thank the IUCAA Reference Centre at North Bengal University for extending necessary research facilities to initiate the work.

\end{document}